# A network analysis of the global energy market: an insight on the entanglement between crude oil and the world economy

Franco Ruzzenenti[1*], Francesco Picciolo[1], Andreas Papandreou[2]

**1** Department of Biotechnology, Chemistry and Pharmacy, University of Siena, Via Aldo Moro 1, IT-53100. Siena, Italy.
**2** Department of Political Economy, National and Kapodistrian University of Athens, Pesmazoglou str 8, Athens, Greece.

**Abstract**

One major hurdle in the road toward a low carbon economy is the present entanglement of developed economies with oil. This tight relationship is mirrored in the correlation between most of economic indicators with oil price. This paper addresses the role of oil compared to the other three main energy commodities -coal, gas and electricity, in shaping the international trading network (ITW or WTW, world trade web) in the light of network theory. It initially surveys briefly the literature on the correlation between oil prices with economic growth and compares the concepts of time correlation with the concept of spatial correlation brought about by network theory. It outlines the conceptual framework underpinning the network measures adopted in the analysis and results are presented. Three measures are taken into account: the ratio of mutual exchanges in the network (reciprocity); the role of distances in determining trades (spatial filling); and the spatial correlation of energy commodities with the whole trade network and with four trading categories: food, capital goods, intermediate goods and consumption goods. The analysis deliver five main results:1) the the energy commodities network was structurally stable amid dramatic growth during the decade considered;  2) that oil is the most correlated energy commodity to the world trade web; 3) that oil is the most pervasive network, though coal is the less affected by distances; 4) that oil has a remarkably high level of internal reciprocity and external overlapping 5) that the reciprocity of the trade network is negative correlated in time with the price of oil.

*corresponding author: ruzzenenti@gmail.com

## 1. Introduction

The way toward a low carbon economy finds one major hurdle in the dependence of developed economies to fossil fuels. Among fossil fuels, oil is generally considered a special form of energy, entangled to the economic process more than any other energy commodities. Indeed, many economic indicators, like the Gross Domestic Product, the Industrial Production or inflation rate, are historically strictly -negative or positive, correlated to oil price. Why oil is so important compared to other fossil fuels? Is there a *physical* basis to correlation of oil price to economy or is it due to monetary and financial reasons? Indeed, oil enters the production function in several sectors of the economy not only as an energy source. Oil is gasoline, plastic, chemicals and finance (Ruzzenenti, 2015a). This paper addresses these questions in the light of network theory by developing a comparative analysis, with a spectrum of network indicators, of the four main energy commodities in relation to the structure of global trade. International trade after globalization became a prominent sector in world economy, whose affluence is now strictly connected to economic growth and development. Furthermore, international trade heavily relies on the transport system, a sector that is almost entirely dependent on fossil fuels and oil particularly. Hence, the World Trade Web (WTW) provides us with an interesting and informative case-study to investigated the role of oil and fossil fuel prices in shaping the world economy.

### 1.1 Temporal correlation and spatial analysis in energy markets, state of art

The price of oil is customarily taken as benchmark for the price of energy. Indeed many price indices in the energy market are locked to the oil price (electricity, natural gas). Moreover, this linking (or coupling) is widely considered a major factor contributing to inflation. Nonetheless, it is less widely understood why oil plays such a pivotal role in the world economy. It is commonly believed that this supremacy is due to the fact that oil is the main energy source for world economy, yet coal as a primary energy source has almost reached oil in share. In 2014 coal energy consumption worldwide was 30% and oil 32% (BP, 2015).

Coal was the fundamental energy source centuries ago that used to shape the economy and to determine the rise and fall of nations (Jevons, 1866). Nowadays oil is indisputably dominating the world economy. Data from the second world war onward show that economic growth is intimately linked to the price of oil (Hamilton, 1983). The issue of correlation, and thereby causality, between the price of oil on one hand and several economic indicators on the other hand, has always been debated in scientific literature (Hooker, 1996; Hamilton, 1996; Papapetrou, 2001; He et al., 2010). Although historic trends show that economy and oil prices are entangled, the extent, the time lag, and the scope of this relationship is still an open issue. However, it appears that the scientific community has converged towards some key points on the topic: 1) the correlation seems to be asymmetric, especially in the short run: when oil prices rise the economy slows down, whereas the opposite relationship is more relaxed (Huang et al. 2005; Rahman and Serletis, 2010, Alvarez-Ramirez et al, 2010); 2) the correlation between oil prices and world economy has diminished, reaching a peak during the oil crisis and weakening in the aftermath (Thang et al., 2010; Naccache, 2010; Alvarez-Ramirez et al, 2010); 3) the correlation increases during oil shocks, indicating that spikes of oil prices are the most influential in affecting the economy (Hamilton, 2003; Hsu et al, 2008; Cologni and Manera, 2009; Jamazzi, 2012); 4) the causality relationship between oil prices and economic growth, measured by statistical means, is ambiguous and sometimes reversed, depending on the country selected or the time period (Narayan and Smyth, 2007; Cologni and Manera, 2009; Benhmand, 2012, Ratti et al. 2013).

We want here to seek a different perspective over the interdependence of energy and economy that goes beyond the customary temporal approach and pursues a spatial approach based on network theory. The network structure of the productive space will be analysed in order to establish a structural correlation between commodities, class of commodities and energy commodities.

The spatial structure of the energy markets has already been the object of studies. So far, it has received the attention of scholars to address the issue of energy security, pointing at the spatial distribution of energy supply chain (Bert Kruyt et al. 2009; Cohen, G et al., 2011; Winzer C., 2012, Zhang, H.-Y., et al., 2015). More recently, network theory has been applied to the global oil market with the aim of analysing its structure, stability and the role and position of countries in international trade. (Fan et al, 2014; Mileva, and Siegfried, 2012; Ji, Q et al., 2014). Network theory has also been used in the field of energy markets to investigate the hierarchical structure and relationship between crude prices worldwide (Zhang, H.-Y., et al., 2014). The goal of this work is that of comparing the four main energy commodities in relation to the structure of trade globally. The remaining of the article is structured as follows: in the next section we will introduce the concept of spatial correlation departing from the Pearson correlation index. We will then develop a binary network analysis of the structure of the global energy market and a weighted analysis of flows, in monetary and mass units. In section 2.4 will recall the Pearson index to investigate the relationship of the four energy commodities with global trade and four trading categories. In the following section will integrate temporal and spatial analysis by comparing oil price with the reciprocity in global trade (symmetry of matrix of trade). In the last section some concluding remarks will be outlined.

## 2.1 From temporal correlation to spatial correlation

In statistics, the Pearson product-moment correlation coefficient is a measure of the linear correlation

(dependence) between two variables C and C', where $p_t^C$ and $\acute{p}_T^c$ are the price (or log return) at time $t$ and the average price in the time interval $T$ of the good C. The Pearson index ranges between +1 and −1 inclusive, where 1 is total positive correlation, 0 is no correlation, and −1 is total negative correlation:

$$\varphi_t^{cc'} = \frac{\sum_{t=1}^T [p_t^c - \acute{p}_T^c] * [p_t^{c'} - \acute{p}_T^{c'}]}{\sqrt{\sum_{t=1}^T [p_t^c - \acute{p}_T^c]^2 \sum_{t=1}^T [p_t^{c'} - \acute{p}_T^{c'}]^2}} \quad (1)$$

In the Pearson corr. index, the space is one-dimensional and the summation runs over the time-variable. In a Network with N nodes, we have N(N-1) possible connections (links) and if the network exchanges two types of flows/commodities, they can be computed with the Pearson corr. index according to the sequence of nodes' links. In this case, the summation runs over the couple of nodes' indices i and j, fro the two flows c and c' :

$$\varphi_w^{cc'}(t) = \frac{\sum_{i \neq j} [w_{ij,t}^c - \acute{w}_t^c] * [w_{ij,t}^{c'} - \acute{w}_t^{c'}]}{\sqrt{\sum_{t=1}^T [w_{ij,t}^c - \acute{w}_t^c]^2 \sum_{t=1}^T [w_{ij,t}^{c'} - \acute{w}_t^{c'}]^2}} \quad (2)$$

This diverse computation of Pearson index expresses a *spatial* correlation because it measures the topological relationship between every couple of links' weight in two networks that share the same nodes (Barighozzi et al., 2010; Ruzzenenti et al. 2015). In other words, the exogenous variable it is not time but topology. We will now investigate the spatial correlation in the productive space, namely, the international trade network of commodities.

The conceptual framework behind network analysis of the productive space is the same as that behind the "too-interconnected-to-fail" theory (in place of the "too-big-to-fail") that places greater attention to the most interconnected nodes and to their role in spreading contagion, rather than to the merely largest ones.

In the following, we will compare four major energy commodities in the framework of the world international trading network (world trade web) through the lens of complex network theory. Henceforth, a country will be considered as a node or vertex of a graph and a trading relationship as a link or edge. Links might have different weights according to trading volumes and the incoming or outgoing vertex might be of different size according to the GDP of the country's economy. In the world economy there are fossil fuel producers and fossil fuel consumers (in red and in blue in Fig. 1). Energy-reach countries exchange energy commodities for consumption or capital goods from energy-poor countries. The structure of this exchange relationship will be the object of the present investigation.

Arguably, the dependency of the world economy on an energy source depends, inter alia, on the degree of interconnection and its capillarity. Thus, the primary information we should look for when considering the world network of energy commodities is: what is the most interconnected and extended energy source?

## 2.1 Spatial correlation: binary analysis

The first step of the analysis required to answer this question concerns the binary structure of the network. A binary, directed graph is specified by a N×N adjacency matrix, A, where N is the number of nodes and the generic entry $a_{ij}$ is 1 when there is a connection from node i to node j, and 0 otherwise. In order to assess the degree of *spreading* of a network in its embedding space, we need to develop a measure that includes both spatial constrains –distances, and topological constrains - structure (Squartini et al., 2013). The simplest definition of a global measure incorporating distances and network structure is

$$F = \sum_{i=1}^{N} \sum_{j \neq i}^{N} a_{ij} d_{ij} \quad (3)$$

where $d_{ij}$ is the generic entry of the matrix of distances, D, among nodes (Ruzzenenti et al. 2012). Since we will consider networks without self-loops (i.e. $a_{ii}=0$), F is a measure of the total distance between different, topologically connected pairs of nodes. Equivalently F can be seen as a measure of the extent to which networks "fill" the available space or, in other words, a measure of the "pervasiveness" of the network.

The quantity F reaches its minimum when the links are placed between the closest vertices. Formally speaking, if we consider the list of all non-diagonal elements of D ordered from the smallest to the largest $d_n^\uparrow \leq d_{n+1}^\uparrow$ , the minimum value of F is simply given by $F_{min} = \sum_{n=1}^{L} d_n^\uparrow$ , where $L = \sum_{i=1}^{N} \sum_{j=1}^{N} a_{ij}$ is the number of links in the network. Similarly, the maximum value of F is reached when links are placed between the spatially farthest nodes. Considering the list $V^\downarrow = (d_1^\downarrow, \cdots, d_n^\downarrow, \cdots, d_{N(N-1)}^\downarrow)$ of distances in decreasing order $d_n^\downarrow \geq d_{n+1}^\downarrow$, the maximum value of F for a network with L vertices is $F_{max} = \sum_{n=1}^{L} d_n^\downarrow$.

In order to compare, and possibly rank, different networks according to their values of F, a normalized quantity should be used. An improved global definition, which we will denote as filling coefficient, is

$$f = \frac{F - F_{min}}{F_{max} - F_{min}} = \frac{\sum_{i=1}^{N}\sum_{j\neq i} a_{ij} d_{ij} - F_{min}}{F_{max} - F_{min}} \quad (4)$$

It is noteworthy that in dense networks with a broad degree (the sum of ingoing and outgoing links) distribution, topology may affect spatial interactions (Squartini et al, 2013a). For example: two highly connected nodes (hubs), as two big economies, are more likely to interact regardless of their physical distance and, if we aim at assessing the role of distances, it is best to disentangle spatial effects from non spatial effects. A way to do that is by adopting a null model (NM), based on the statistics of exponential random graphs, to compare and weight our measures (Squartini et al, 2013a). Null models are characterized by some kind of topological property (such as the link density or the degree sequence) that is *a priori* independent of any spatial constraint. Through a NM we can compute the probability $p_{ij}^{NM}$ that is the expected value given the chosen constraint. By means of NM we can improve our definition of filling coefficient by filtering out the spurious spatial effects due to the non-spatial constraint enforced. In order to achieve this result, a comparison between the observed value of f and its expectation is needed. Consider the expected value of the filling coefficient under any of the three aforementioned null models (NM):

$$\langle f \rangle_{NM} = \frac{\sum_{i=1}^{N}\sum_{j\neq i} p_{ij}^{NM} d_{ij} - F_{min}}{F_{max} - F_{min}} \quad (5)$$

where $p_{ij}^{NM}$ is given by the chosen null model. The comparison between observation and expectation can be easily carried out by making use of the following rescaled version of the filling coefficient that we denote as filtered filling (Ruzzenenti et al. 2012):

$$\varphi_{NM} \equiv \frac{f - \langle f \rangle_{NM}}{1 - \langle f \rangle_{NM}} \quad (6)$$

Hence, φ (equation 6) and *f* (equation 4) carry out two different information about the spatial structure of the network. The question addressed by φ is: *how much the network is affected by distance?* Whereas the question addressed by *f* is: *how much the network is stretched in its embedding space?*

Besides space filling, an important feature of networks is *reciprocity*. Indeed, for the present analysis reciprocity is of paramount interest: Are energy source producers more or less reciprocated in the world differentiated economic system by producers of goods and capital?

To answer this question, we consider the definition of reciprocity for a binary network

$$r = \frac{L^{\leftrightarrow}}{L} = \frac{\sum_{i=1}^{N}\sum_{j\neq i} a_{ij} a_{ji}}{\sum_{i=1}^{N}\sum_{j\neq i} a_{ij}} \quad (7)$$

where $L^{\leftrightarrow}$ is the number of reciprocated links (going both ways between pairs of vertices), and L is the total number of links (Newman Forrest and Balthrop 2002). Exactly as in the case of the filling coefficient, the binary reciprocity has a "filtered" counterpart, defined in the same way (Garlaschelli and Loffredo 2004, Ruzzenenti et al., 2010), incorporating both the observed and the expected values under a chosen null model (NM):

$$\rho_{NM} = \frac{r - \langle r \rangle_{NM}}{1 - \langle r \rangle_{NM}} \quad (8)$$

Results of the binary analysis are reported on Table 1. The data set employed for the analysis is BACI, developed by CEPII[1] and covers a decade from 1998 to 2007. in order to avoid anomalies due to the crisis, the scope of the analysis ends before 2008. The first notable information delivered by the binary analysis is that the structure of the energy network is stable. Although both the number of links and the trading volume of all the energy commodities considered (except electricity) grew dramatically between 1998-2007, significantly more than global trade, the spatial and symmetrical structure of the four energy markets is pretty stable. Oil grew 554% in value, 56% in mass and 57% in number of links. Coal grew 188%, 82% and 37% respectively and gas 295%, 172% and 46% whereas the world trade network grew 158% in value, 51% in mass and 33% in number of links (Table 2 and Table 3). However, if we look at the standard deviation for all the selected measures, the low values indicate that the network is structurally stable (Table 1). This means that the binary reciprocity, the connectance (density of links) and the spatial filling of energy commodities did not significantly changed in the decade considered. More in details, oil is by far the most *connected* energy commodity, with, on average, 1 out of 10 possible links present in the network, whereas electricity is the least connected, with a connectance of 0.02 on average. The spatial analysis indicate that crude oil and second coal present the highest spreading, perhaps not surprisingly, given that the transport of gas and electricity is bound to costly and complex infrastructures. Interestingly, coal is the energy commodity less affected by distances in transports, as the highest φ value indicate. This is probably due to the fact that solid energy sources are more suitable for transport than liquid energy sources. As previously highlighted, oil is a denser network and it is also much more reciprocated, as indicated by *r*. However, as the ρ value shows, *it is not more reciprocated because it is denser*. Therefore, we must deduce that countries that exports oil are much more likely to establish mutual links than other energy-commodities exporting countries.

[1] http://www.cepii.fr/CEPII/en/bdd_modele/presentation.asp?id=1

## 2.3 Spatial correlation: weighted analysis

However, the binary reciprocity measure is not enlightening on the *extent* of the mutual relationship. A weighted reciprocity measure is needed in order to assess the amount of reciprocal flow. Furthermore, switching from the binary analysis to the weighted description of networks enable us to evaluate trades both in monetary terms and in mass flows. A weighted, directed graph is specified by a N×N adjacency matrix, W, where N is the number of nodes and the generic entry $w_{ij}$ is the intensity of the connection (the amount of trade flow for WTW) from node *i* to node *j*. It is noteworthy that for every couple of interacting nodes, the reciprocated part of two counteractive flows is the *minimum* flow in between (Squartini et al 2013b):

$$w_{ij}^{\leftrightarrow} = min[w_{ij}, w_{ji}] = w_{ji}^{\leftrightarrow} \quad (9)$$

Therefore, analogously to equation (7), a normalized measure of the weighted reciprocity of a network can be defined as follow:

$$r^w = \frac{W^{\leftrightarrow}}{W} = \frac{\sum_{i=1}^{N}\sum_{j\neq i} w_{ij}^{\leftrightarrow}}{\sum_{i=1}^{N}\sum_{j\neq i} w_{ij}} \quad (10)$$

Likewise, it is possible to define a rescaled measure of the reciprocated flow in two layer, layers A and B, of a multiplex as follows interacting networks, for every couple of nodes:

$$w_{ij}^{\leftrightarrow}(AB)_{syn} = min\left[\frac{w_{ij}^A}{W^A}, \frac{w_{ij}^B}{W^B}\right] \quad (11a)$$

$$w_{ji}^{\leftrightarrow}(AB)_{rev} = min\left[\frac{w_{ij}^A}{W^A}, \frac{w_{ji}^B}{W^B}\right] w_{ji}^{\leftrightarrow}(AB)_{rev} \quad (11b)$$

And a measure of the overlapping *synergic* (*reverse*) flows going in the same (opposite) directions within the two networks:

$$C^{AB}_{syn(rev)} = \frac{W^{\leftrightarrow}_{syn(rev)}}{W} = \sum_{i=1}^{N}\sum_{j\geq i} w_{ij}^{\leftrightarrow}(AB)_{syn(rev)} \quad (12)$$

This latter measure is intended to indicate the percentage of the import/export that two networks share for every couple of nodes (countries). For example, a 25% reverse overlapping index between oil and world trade web, means that 25% of exports in the oil market is reciprocated by the world economy directly from the importing country (it could be interpreted as a *reinvestment* index or, for synergic flows, as a *complementary* index).

Table 2 shows results for the monetary and mass flows for the years 1998 and 2007. Despite the fact that world consumption of coal matched oil consumption, oil as a commodity, still represents the greatest energy commodity market in volume, by one order of magnitude. But not in mass: in 2007 oil and coal have been trade for almost the same amount. Nevertheless, the network analysis shows that oil stems out for its structural role compared to the other energy commodities. The first striking result is weighted reciprocity: oil scored in 1998 and 18% and 20% in 2007 of reciprocity. This means that one fifth of oil is mutually traded in the world. This is much higher, for one order of magnitude, compared to the other energy commodities. Indeed, oil exporters countries are also oil importers. The oil network scored in 2007 also the highest overlapping index in synergic flows and reverse flows. Oil exchange directly (reverse flows from the importing country) around 35% (34% in 1998) in value and drives complementary flows (toward the importing country) for 40% (37% in 1998) of exports in value. The results hold for the mass flows, albeit the physical reciprocity and the overlapping indices are much lower for all the networks considered (Table 3).

**2.4 Spatial correlation of the energy commodities with the disaggregated economic network**

The last part of our analysis will focus on the spatial correlation (equation 2) of the four energy commodities with the WTW and with four major categories of trading goods: food, capital goods, intermediate goods and consumption goods (Table 4). It is noteworthy that the energy commodities are classified either as intermediate, like oil and in some cases coal and gas, or as consumption goods, like electricity. The correlation measures are averaged over 10 years, from 1998 to 2007. Surprisingly, the most correlated energy commodity to the economy electricity, with the exception of oil with food (Table 5). On the other hand, intermediate goods are in general the most correlated category with the energy commodities. However, if we swop the directions of flows and consider only reverse flows, that is, $w_{ij}^A$ with $w_{ji}^B$ instead of $w_{ij}^B$, we get a different picture. While intermediate goods are still the most correlated category of goods with the energy sector, oil now is the most correlated energy economy with the economy (Table 6).

Based on the Pearson correlation index (equation 2), it is possible to develop measure of the distance between every couple of networks (Mantegna, 1999):

$$d_{w/u}^{c,c'} = \sqrt{\frac{1 - \varphi_{w/u}^{c,c'}}{2}} \quad (13)$$

The more correlated are two networks, the closer are in distance. We can further project this distance metrics into a *dendrogram*, to depict the degree of relationship among networks. A dendrogram is a hierarchical tree diagram used, mainly in genetics (phylogenetic tree), to portray the hierarchy of the inferred evolutionary relationships among various biological species. In our case it shows the hierarchy of clusters and a cluster is defined by a group of tightly correlated commodities. Figure 2 shows the temperature map of the correlation matrix and the dendrogram of the Pearson correlation index for flows going in the same direction (synergic flows). Clusters are visible on the matrix as blocks of homogenous intensity. Gas and electricity form a cluster and they are loosely connected with oil, whereas coal is set apart, being unrelated with the other energy commodities but also with the economy. The dendrogram on the right of Figure 3 shows this structure of relationship, where coal branches on the top of the tree and oil, gas and electricity are more closely related to the economic cluster. In Figure 3 we show temperature map of the correlation matrix of the *reverse* flows and the dendrogram. The dendrogram can deliver the information about the hierarchical structure of the clusters. In the second dendrogram oil is tightly related to the economic cluster and the other three energy commodities branch earlier in the tree. This result indicates that the topology of the oil market is sharply different from the topology of the other energy commodities, which are in relation to the world trade mostly synergistically. On the contrary, oil market excahnges in return from the economy, and prominently from the intermediate goods market, more than any other energy markets.

## 2.5 Integrating spatial and temporal analysis

We have demonstrated the centrality in the world trading network of oil compared to other energy sources by analysing the spatial (topological and metric space) correlation among networks. It is of our interest now to investigate the temporal correlation between the network representation of world economy and the price of oil. As we have previously highlighted, weighted reciprocity (equation 10) is an insightful measure of networks. Figure 4 shows the historic trend of reciprocity (red line) compared to oil price (constant price, base year 2011, source BP), from 1960 to 2010. The Pearson correlation index of the two curves is -0.70 (Picciolo et al., 2015).

However, as it is shown in Table 2, the world trade web is highly reciprocated, meaning that on average $w_{ij}$ is equal to $w_{ji}$: the matrix is almost symmetrical. In networks with broadly distributed strengths (total export and total import) the attainable level of symmetry strongly depends on the in- and out-strengths of the end-point vertices: unless they are perfectly balanced, it becomes more and more difficult, as the heterogeneity of strengths across vertices increases, to match all the constraints required to ensure that $w_{ij} = w_{ji}$ for all pairs. Therefore, even networks that maximize the level of reciprocity, given the values of the strengths of all vertices, are in general not symmetric. Moreover, in networks with balance of flows at the vertex level (export ~import for all vertices) an average symmetry of weights is automatically achieved by pure chance, even without introducing a tendency to reciprocate. In many real networks, the balance of flows at the vertex level is actually realized, either exactly or approximately, as the result of conservation laws (i.e. balance of payments). In those cases, the symmetry of weights should not be interpreted as a preference for reciprocated interactions. We have thus developed a filtered measure of reciprocity, formally similar to the $\rho_{NM}$ in equation 8, that encapsulated a null model aimed at imposing the vertex balancing constrain. In what follows, the reciprocity adjusted measure discounts the effect of the tendency of nodes to preserve the balance between import and exports (balance of payments). Fig 4 shows that the adjusted reciprocity (green line) heightens the correlation with oil price (correlation index: -0.78) but down shifts the curve, indicating that the network, washed by the balancing effect, is less reciprocated. This is noteworthy, as it means that the correlation between oil and reciprocity does not depend from the global imbalance (a purely monetary effect), but it depends on some structural reason that links the way countries reciprocate their trade to the price of oil. This reason could lay in the development of the global value chain and how it depends on the transport system (Picciolo et al, 2015).

Nevertheless, the most interesting information that comes from comparing oil price and reciprocity, rests in the temporal sequence of curve peaks. Oil curves, more clearly during the first oil crisis peak after and not before the reciprocity curve, as is clearer in the small frame that shows a magnification of the two curves (Fig 4).

Indeed, the reciprocity starts declining three years before the oil shock. In spite of existing literature that suggests that the oil shock of the 1973 followed instead of preceding the economic recession, it is still widely believed that the oil crisis caused the economic crisis rather than the opposite (Cologni and Manera, 2009; Ratti and Vespignani, 2013). Although several economic indicators seem to follow the trend of oil peaks, it is clear by looking at trends in industrial production or retail sales, that none of these seem to display such a clear anticipation of the oil peak in the 1970s as does network reciprocity (Figure 5).

However, the economic interpretation of reciprocity is still uncertain. Arguably reciprocity is linked to economic recession. One reason for this relationship could be the pressure exerted on balance of payments by recession and high energy prices. A global node unbalancing may cause a symmetry breaking in the network. Nonetheless, as was previously explained, the adjusted reciprocity should discount the effect of node unbalancing but $\rho_{NM}$ emphasizes the reciprocity trend, rather than levelling it. Henceforth, the high correlation between $\rho_{NM}$ (the adjusted reciprocity) and oil price informs us that, beyond commercial balance effect, lower oil prices stimulate reciprocal trades.

It is noteworthy that the trend in reciprocity after the first oil crisis, though being still so correlated to oil prices, it seems not to anticipate oil spikes. Evidently, the first oil crisis should be considered a remarkable exception in the fundamental trend interlinking oil price and recession. It is likely that recession in the 1970s came first and the oil crisis followed, but the same it can't be said for the following oil shocks.

## 3  Concluding remarks

In this work we have developed a network analysis of the global oil market compared to other three energy commodities in relation to the world trade network, aggregate and disaggregate to four main categories: food, capital goods, intermediate goods and consumption goods. The goal of the present analysis is that of shedding a new light on the entanglement of oil with the world economy by means of measures derived from network theory. The idea is that spatial correlation can provide a different information from time correlation analysis and open new lands of investigation for further research. Results show that oil is the most interconnected and pervasive energy commodity inasmuch as it has the highest filling value and the highest connectance. Perhaps, this is not surprisingly, given the notion that oil is the first energy source in the world. What is more surprising is that oil is also the most reciprocated energy commodity in the world, both in binary and weighted terms. This result seems to corroborate the hypothesis that oil is a mature and complex market, where exporters are also importers. The high level of internal reciprocity, for example, is typical of commodity markets positioned higher in the value chain (Gemmetto et al, 2015). This hypothesis is further supported by the spatial correlation analysis comparing the energy commodities with global trade. Oil is in general, though slightly, more correlated to the trading categories considered. When assessing reverse flows, the gap in correlation with the other energy commodities widens. This result suggests that the interdependence of oil with economy, and prominently with intermediate goods, unfolds when considering flows in opposite directions. Oil exchanges more with the rest of the global economy than any other energy commodity. This higher purchasing power of oil is further confirmed by observing the overlapping index. All these results together hint to the structural role of oil which is unmatched by any other energy source. In the last part of the paper we show that the structural role of oil roots back in time and that since the 1960s the negative correlation of the symmetry of the matrix of world trade with oil is remarkably high. We proved that this tight correlation is not due to monetary effects –the balance of payments, but to more profound and structural reasons. This result tells us that mutual trade has always been connected to the price of oil, rather than trading agreements or differential in labour costs and probably hints to the not-fungible role of oil in the transport sector, which shaped the production chain globally. Nevertheless, more research is needed in order to understand the longstanding nexus between oil price and the global value chain. Albeit only by a preliminary and general network analysis, mainly aimed at providing an overview of some analytical tools applicable at energy markets, we have here shed a new light into the structural relationship between oil and economy. The authors of this article believe that the degree, the scope and the causes of the entanglement of economy with oil it is still little understood. Despite the copious literature on the subject and an unanimous consensus on the leading role of oil in the present economy, oil is still regarded by most scholars and policy actors as a mere energy source. On the track toward a carbon-free economy, can we keep considering oil just as another energy source and replaceable as long as production function and the relative price allows it? Oil shapes the present, global economy in many ways, ranging from material production to transports, from chemistry to agriculture, from the monetary system to the financial sector. Our world is composed by oil derivatives, directly or indirectly. Oil is transfused into the production of most things, in so far that it resembles what for some ancient Greek philosophers was the *apeiron*. The unmoulded, unlimited, primal substance that yields the building material for the World. Oil, we must acknowledged, is not unlimited though.

*Table 1 Binary network analysis of the energy market, mean values and standard deviation over the period 1998-2007.*

| 1998-2007 | Reciprocity | Connectance (link density) | Spatial filling (f) | Rho | Phi |
|---|---|---|---|---|---|
| Oil | 0.52±0.008 | 0.10-0.16 | 0.25±0.003 | 0.21±0.008 | -0.17±0.007 |
| Coal | 0.36±0.01 | 0.04-0.05 | 0.23±0.004 | 0.12±0.01 | -0.14±0.008 |
| Gas | 0.33±0.01 | 0.03-0.04 | 0.20±0.007 | 0.17±0.01 | -0.21±0.01 |
| Electricity | 0.19±0.02 | 0.01-0.02 | 0.18±0.005 | 0.12±0.01 | -0.29±0.01 |
| WTW | 0.84* | 0.56* | 0.34* | 0.21* | -0.10* |
| Food | 0.68* | 0.28* | 0.32* | 0.17* | -0.13* |

*values refer to year 2007

*Table 2 Weighted network analysis of energy market, values in thousand dollars, in 2007 and 1998.*

| Year: 2007 (1998) | Reciprocity | Overlapping Index: *syn* flows | Overlapping Index: *rev* flows | W tot | Growth rate | links | Growth rate |
|---|---|---|---|---|---|---|---|
| WTW | 0.69 (0.76) | | | 1.24*10^10 (4.8*10^9) | 158% | 2.4*10^4 (1.8*10^4) | 33% |
| Oil | 0.25 (0.21) | 0.40 (0.37) | 0.34 (0.35) | 5.3*10^8 (8.1*10^7) | 554% | 6.6*10^3 (4.2*10^3) | 57% |
| Coal | 0.07 (0.11) | 0.22 (0.26) | 0.20 (0.24) | 7.2*10^7 (2.5*10^7) | 188% | 2.2*10^3 1.6*10^3 | 37% |
| Gas | 0.08 (0.03) | 0.21 (0.16) | 0.17 (0.14) | 1.5*10^8 (3.8*10^7) | 295% | 1.9*10^3 (1.3*10^3) | 46% |
| Electricity | 0.09 (0.12) | 0.20 (0.25) | 0.19 (0.24) | 2.5*10^5 (1.48*10^5) | 79% | 7.6*10^2 5.8*10^2 | 31% |

*Table 3 Weighted network analysis of energy market, values in tons, in 2007 and 1998.*

| Year: 2007 (1998) | Reciprocity | Overlapping Index: *syn* flows | Overlapping Index: *rev* flows | W tot (total volume) | Growth |
|---|---|---|---|---|---|
| WTW (ton) | 0.40 (0.48) | | | $1.1*10^{10}$ ($7.3*10^{9}$) | 51% |
| Oil (ton) | 0.20 (0.18) | 0.43 (0.39) | 0.26 (0.26) | $1.0*10^{9}$ ($6.4*10^{8}$) | 56% |
| Coal (ton) | 0.04 (0.06) | 0.26 (0.25) | 0.13 (0.16) | $1.0*10^{9}$ ($5.5*10^{8}$) | 82% |
| Gas ($m^3$) | 0.07 (0.03) | 0.24 (0.23) | 0.11 (0.10) | $7.9*10^{8}$ ($2.9*10^{8}$) | 172% |
| Electricity (kWh) | 0.06 (0.12) | 0.14 (0.16) | 0.11 (0.14) | $8.6*10^{5}$ ($8.4*10^{5}$) | 2% |

*Table 4 List of entries of the correlation matrix: Figures 2 and 3.*

| | |
|---|---|
| 1. | Oil |
| 2. | Coal |
| 3. | Gas |
| 4. | Electricity |
| 5. | Total trade |
| 6. | Food |
| 7. | Capital goods |
| 8. | Intermediate goods |
| 9. | Consumption goods |

*Table 5 Pearson correlation index of synergic flows, spatial correlation, mean values and standard deviation for the period 1998-2007.*

| Synergic flows 1998-2007 | WTW | Food | Capital Goods | Intermediate goods | Consumption goods |
|---|---|---|---|---|---|
| Oil | 0.50±0.04 | 0.48±0.05 | 0.35±0.02 | 0.55±0.05 | 0.37±0.01 |
| Coal | 0.27±0.03 | 0.30±0.04 | 0.22±0.04 | 0.28±0.03 | 0.25±0.04 |
| Natural Gas | 0.50±0.03 | 0.41±0.03 | 0.32±0.04 | 0.56±0.04 | 0.40±0.03 |
| Electricity | 0.53±0.02 | 0.45±0.02 | 0.38±0.05 | 0.58±0.007 | 0.46±0.007 |

*Table 6 Pearson correlation index of reverse flows, spatial correlation, mean values and standard deviation for the period 1998-2007.*

| Reverse flows 1998-2007 | WTW | Food | Capital Goods | Intermediate Goods | Consumption goods |
|---|---|---|---|---|---|
| Oil | 0.43±0.02 | 0.42±0.05 | 0.40±0.02 | 0.44±0.02 | 0.35±0.01 |
| Coal | 0.24±0.04 | 0.17±0.05 | 0.24±0.05 | 0.26±0.04 | 0.19±0.05 |
| Natural Gas | 0.36±0.01 | 0.34±0.04 | 0.34±0.02 | 0.39±0.02 | 0.29±0.02 |
| Electricity | 0.40±0.04 | 0.35±0.03 | 0.37±0.05 | 0.43±0.05 | 0.30±0.03 |

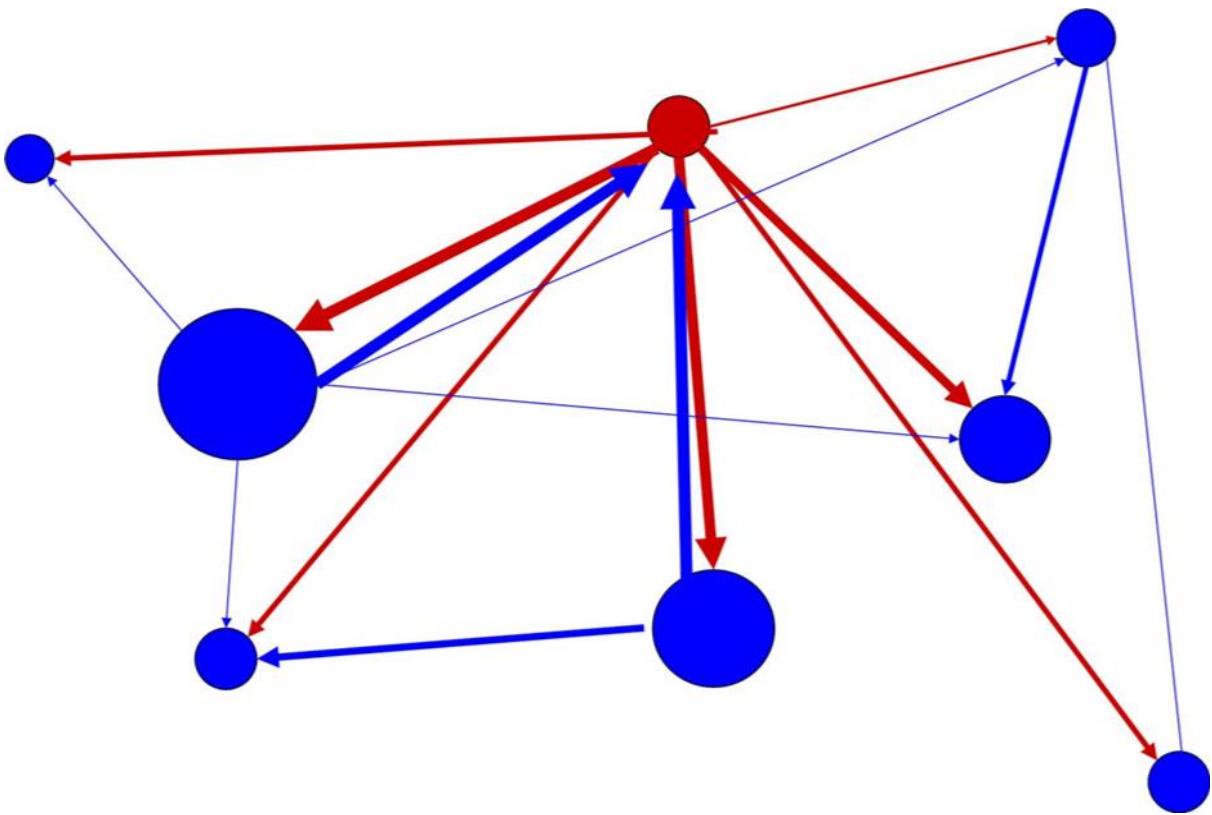

*Figure 1 Spatial correlation: graphic abstract*

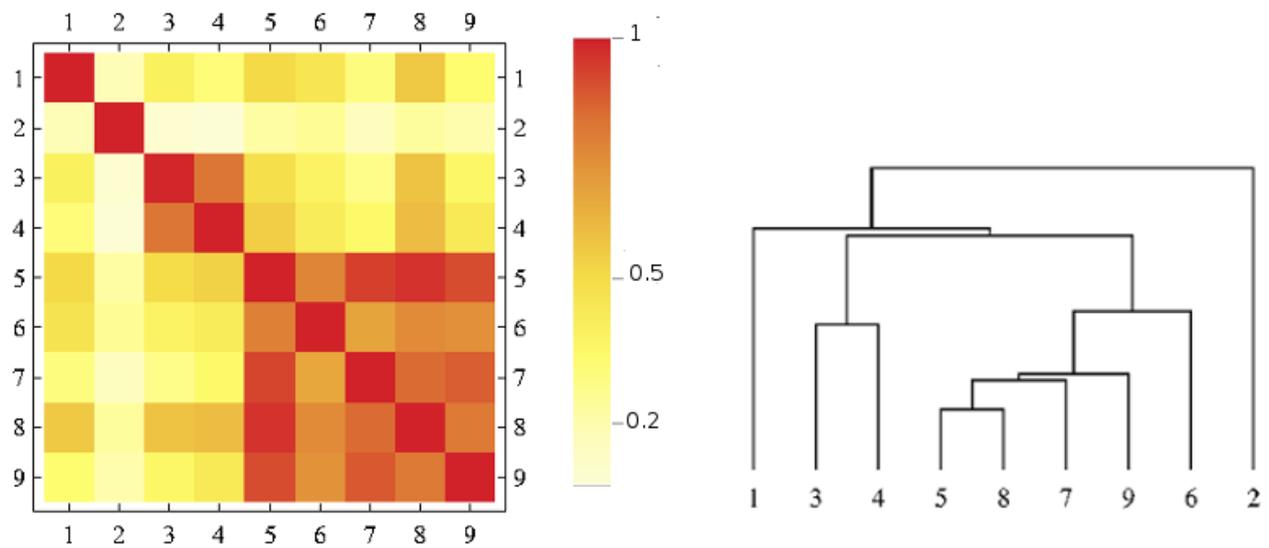

*Figure 2 Pearson correlation index of synergic flows: correlation matrix and dendrogram*

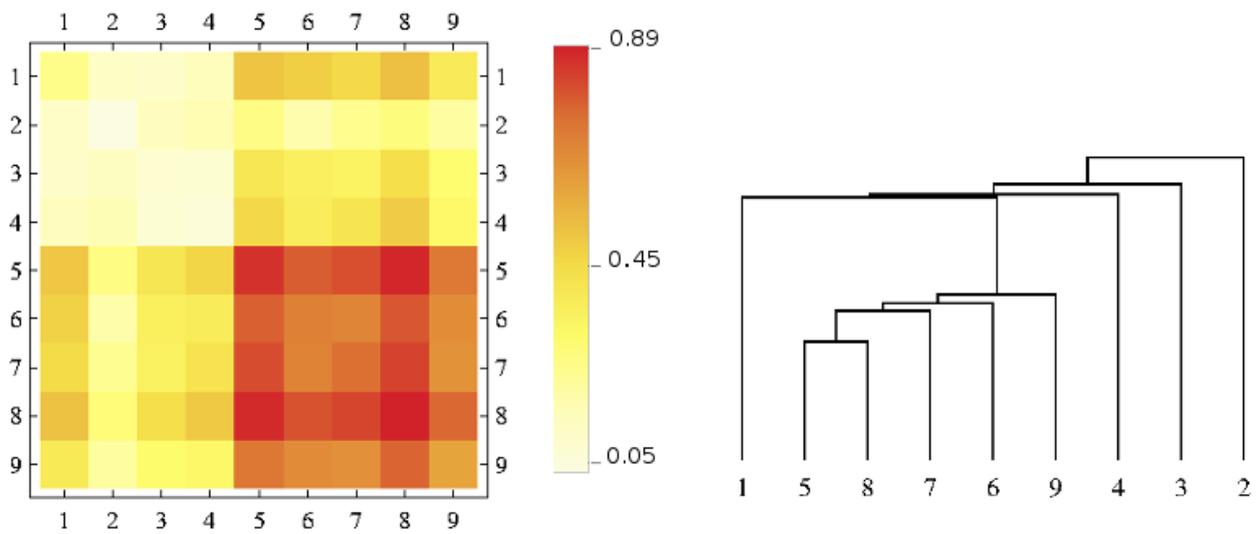

*Figure 3 Pearson correlation index of reverse flows: correlation matrix and dendrogram*

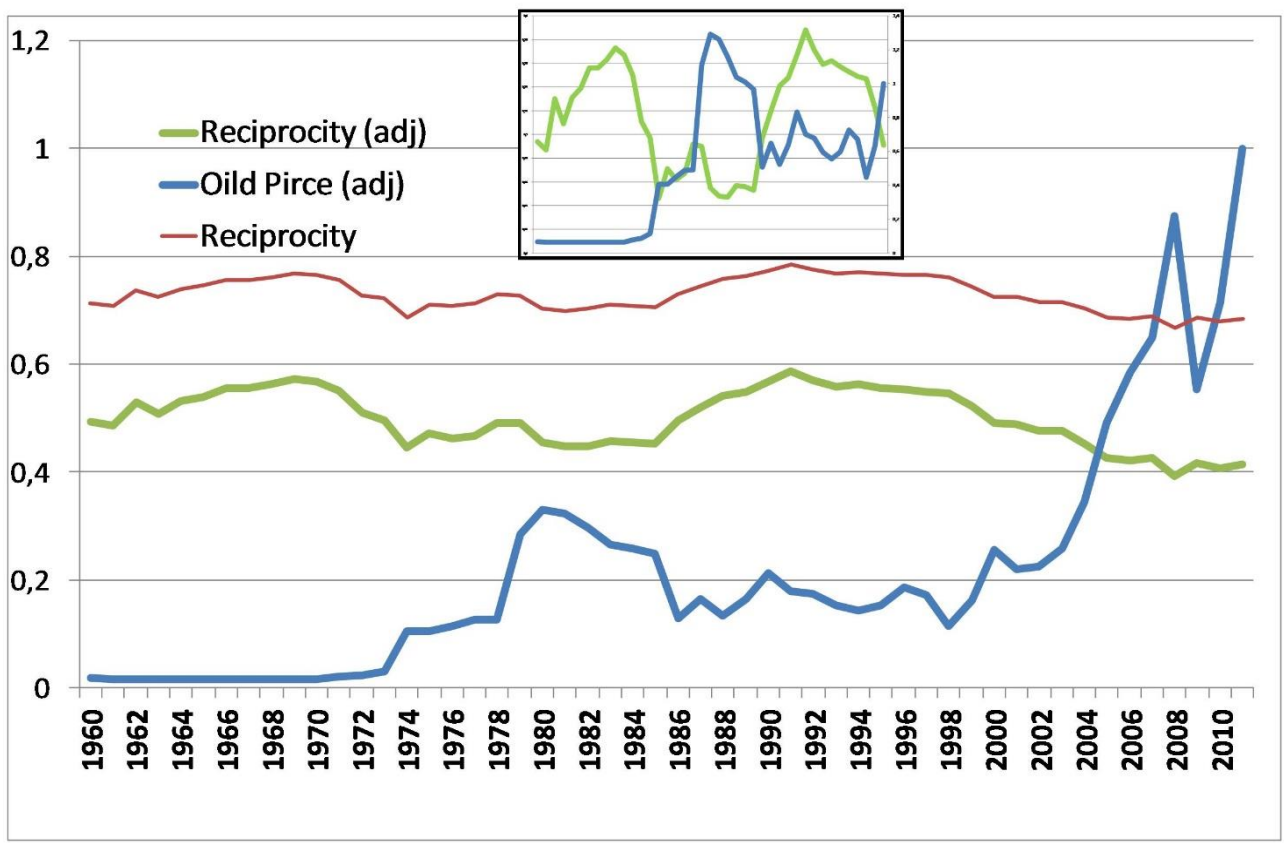

*Figure 4 Correlation between crude oil price, reciprocity and ρ value*

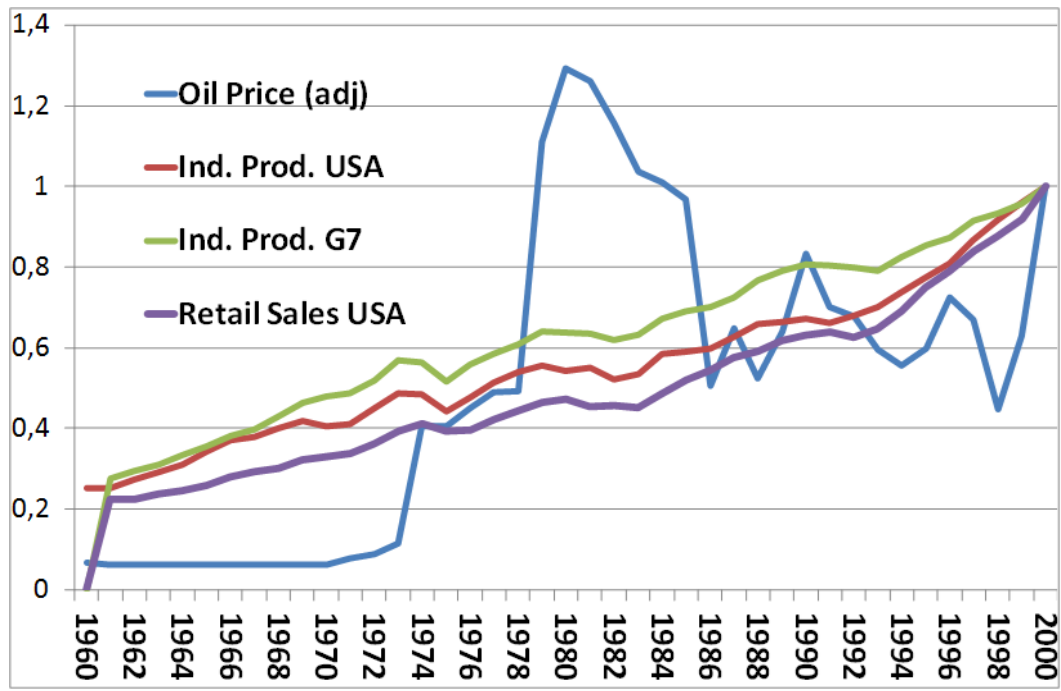

*Figure 5 Trend in industrial production and retail sales compared to oil, base year=2000*